# Individual brain parcellation: Review of methods, validations and applications

Chengyi Li, Shan Yu, Yue Cui, *Member, IEEE*

*Abstract*—**Individual brains vary greatly in morphology, connectivity and organization. The applicability of group-level parcellations is limited by the rapid development of precision medicine today because they do not take into account the variation of parcels at the individual level. Accurate mapping of brain functional regions at the individual level is pivotal for a comprehensive understanding of the variations in brain function and behaviors, early and precise identification of brain abnormalities, as well as personalized treatments for neuropsychiatric disorders. With the development of neuroimaging and machine learning techniques, studies on individual brain parcellation are booming. In this paper, we offer an overview of recent advances in the methodologies of individual brain parcellation, including optimization- and learning-based methods. Comprehensive evaluation metrics to validate individual brain mapping have been introduced. We also review the studies of how individual brain mapping promotes neuroscience research and clinical medicine. Finally, we summarize the major challenges and important future directions of individualized brain parcellation. Collectively, we intend to offer a thorough overview of individual brain parcellation methods, validations, and applications, along with highlighting the current challenges that call for an urgent demand for integrated platforms that integrate datasets, methods, and validations.**

## I. Introduction

The human brain varies greatly in morphology, connectivity and functional organization [5, 6]. Brain atlases, mapping anatomically distinct and functionally specialized brain subregions, are fundamental for understanding complex brain function as well as precise diagnosis and treatments for brain diseases [8]. However, directly registering these group level atlases from standard coordinates to individual space based on morphological information often overlooks inter-subject differences in regional positions and topography, leading to inaccuracies in individual brain mapping and, therefore, fails to capture individual-specific characteristics and cannot guide personalized clinical applications at an individual level [3].

Magnetic resonance imaging (MRI) is a non-invasive technique for measuring brain structure and function, and is essential for obtaining individual-specific brain mapping. Resting-state functional MRI (rsfMRI) has been widely used in individualizing studies. rsfMRI captures the functional activity of the brain during resting states, and is proven to reflect the brain's functional organization [2, 14-18]. Other modalities such as diffusion MRI (dMRI) [4, 11, 19], structural MRI (sMRI) [22], and task-based functional MRI (tfMRI) [3] have also been investigated for individualization. Individual-specific parcellation was first proposed using region growing algorithm based on resting state functional connectivity in 2013 [14]. Since then, various optimization-based approaches were later investigated, including clustering, template matching, graph partition, matrix decomposition and gradient-based methods, characterizing by directly determining individual parcels (i.e., regions) based on predefined assumptions [2, 14-17, 28, 29]. With the advances in neural networks and deep learning techniques, some methods automatedly learn the feature representation of each parcel from the training data, and obtain individual parcellation inferred from the trained model [3, 11, 18]. The individualization techniques have been successfully employed in the field of neuroscience and clinical medicine. For instance, they have been used for interpreting structural and functional variability of the brain and understanding individual-specific brain characteristics such as age [33], sex [36], cognitive behaviors [31], development and aging [33], genetic and environmental factors [38], and translational and comparative research [41]. Individual parcellation also plays a crucial role in identifying biomarkers for various neurological and psychiatric disorders [43-46]. Although individual brain parcellation could promote neuroscience research and personalized clinical solutions, and is a topic of great interest as demonstrated by the rapid development in this field, extensive literature is still limited. To our knowledge, the only published review focuses on precise navigated neuromodulation based on individual brain mapping [49]. Therefore, updated reviews are required to analyze and summarize the actual state of the art.

In this review, we present an overview of recent advances in neuroimaging-based individual brain parcellation, focusing on approaches in terms of optimization-based and learning-based methodologies, optimization objectives (i.e., prior assumptions), and data modalities. Fig. 1 highlights the categories of individual brain parcellation algorithms analyzed in the following section of the paper. We then report on the most commonly adopted metrics for the evaluation of individual brain parcellation results, including intra-subject reliability, intra-parcel homogeneity, capacity of predicting personal characteristics and so on. We also review the progress in individual brain mapping linking brain function and behaviors, identification of biomarkers, precise targeting for

This work was supported by National Science Foundation of China (No. 82371486).

Chengyi Li, Shan Yu, and Yue Cui are with the Laboratory of Brain Atlas and Brain-inspired Intelligence, Institute of Automation, Chinese Academy of Sciences, Beijing, China, and School of Artificial Intelligence, University of Chinese Academy of Sciences, Beijing, China (correspondence e-mail: shan.yu@nlpr.ia.ac.cn; yue.cui@nlpr.ia.ac.cn).

neuromodulation and neurosurgery. The paper concludes with a discussion on future directions and open issues in the field of neuroimaging-based individual brain parcellation.

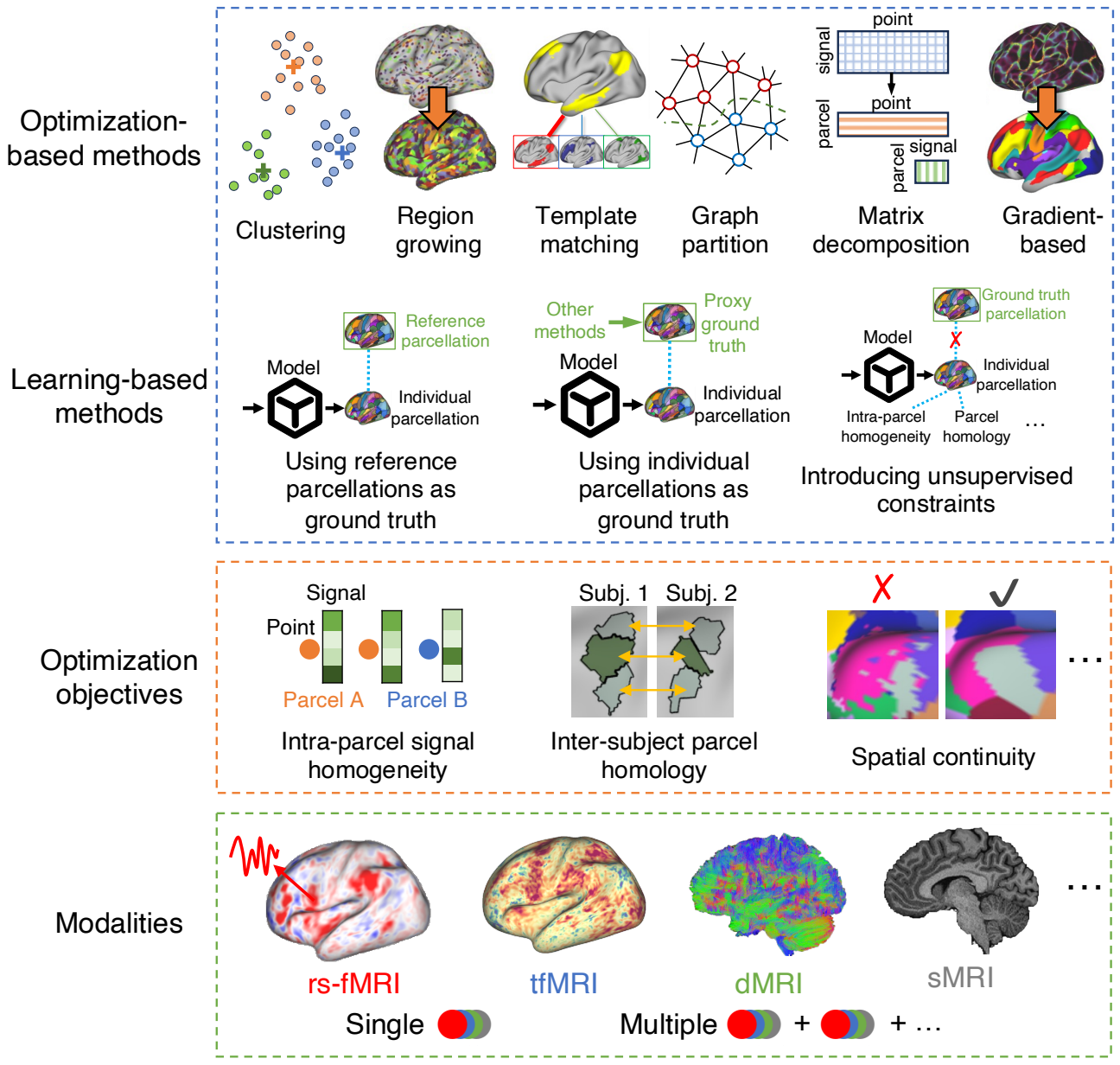

Fig. 1. Categories of individual brain parcellation techniques. These methods are distinct in terms of methodologies including optimization- and learning-based methods, optimization objectives, and data modalities.

## II. Methodologies for Individual Brain Parcellation

Individual brain parcellation is the process of determining the "optimal" assignment of each voxel or vertex in the brain to a parcel (i.e., region) under the guidance of individual information, such as rsfMRI timeseries, anatomical and functional connectivity, and morphological features. In this paper, "individual" is interchangeable with "individualized", "individual-specific" and "personalized". Brain parcellation refers to brain mapping, brain atlases, functional networks or functional regions across the whole brain or cortex in the present paper. The focus is primarily on individualization at the whole-brain/cortex level, and individual parcellation methods that specifically target a fraction of regions, such as the thalamus [53], hippocampus [54], and language areas [56], are not discussed in the present review. Methodologically, individual parcellation techniques can be classified into optimization-based and learning-based methods. Optimization-based methods directly obtain individual parcellations by optimizing predefined assumptions, while learning-based approaches infer individual regions by estimating a parametric model (i.e., mapping) that maps voxel/vertex-wise signals to subject-specific parcellations from data via a training process. Additionally, algorithms vary in terms of optimization objective functions, the use of single or multiple data modalities, and the consideration of single or groups of participants during the individualization procedures.

### A. Optimization-based methods

Optimization-based individual brain parcellation methods allocate voxels/vertices into various parcels mainly based on assumptions that the signals of voxels/vertices within a parcel share similar characteristics. Some approaches also constrain the homology of parcels across subjects and the spatial continuity of each parcel. These assumptions/constraints can be either implicitly embedded within the algorithm or formulated by an explicit predefined objective function. Formally, optimization-based methods can be explained as an optimization process:

$$\underset{\{P_i\}_{i=1}^N}{\operatorname{argmin}} f(\{X_i, P_i\}_{i=1}^N, P_0), \quad (1)$$

where $f(\cdot)$ denotes the predefined objective function, $N$ denotes the number of subjects, $i$ denotes the index of a subject, $X_i$ denotes the individual-specific information, $P_i$ denotes individual parcellation, and $P_0$ denotes a reference parcellation if needed. Some optimization-based methods rely on personal information from a single participant ($N = 1$). In this case, some studies introduced reference parcellation $P_0$ to ensure inter-subject parcel homology by constraining the difference between the individualized and reference parcellation. Other studies take advantage of a group of participants to co-optimize objective functions to ensure the homology of parcels across subjects ($N > 1$). Empirically, the objective function $f(\cdot)$ usually has a simple form, and thus it may not capture high-order and nonlinear correlations between the individual-specific information $X_i$ and the individual parcellation $P_i$. Indeed, the diverse functional complexity of the brain is likely linked to individual-specific characteristics in complex interrelations. We categorized optimization-based methods into six subgroups: clustering, region growing, template matching, graph partition, matrix decomposition, and gradient-based methods (TABLE I).

#### 1) Clustering

Clustering is the task of grouping voxels/vertices that share similar characteristics, ensuring that voxels/vertices in the same parcel (i.e., cluster) are more similar to each other than to those in other parcels. Clustering algorithms adhere to intra-parcel homogeneity and can be employed on single subjects using individual functional, anatomical, and morphological features. However, the direct application of clustering solely considers personal information and lacks regional correspondence at the population level, particularly for fine-scale brain parcellation, limiting comparative analysis across individuals. For example, N-cut clustering has been applied to a local connectivity matrix, which fused resting-state functional connectivity and anatomical connectivity between each voxel and its neighboring voxels, to obtain individual parcellations for single subjects [9, 10]. In addition, simple linear iterative clustering has been used to obtain individual parcellations based on combined functional-spatial distance, in which the spatial distance restricts the search space to the neighborhood of cluster centers [12]. Although both local connectivity and spatial distance ensure the spatial continuity of the parcellation, individuals are not comparable in terms of regional attributes due to the absence of homology. To address this issue, reference parcellations providing group-level prior regional knowledge have been introduced to guarantee inter-subject homology in the case of single-subject clustering [2, 4]. Reference parcellations effectively constrain the topographic similarity of the same brain region across individuals, making individual brain parcellation results more conducive to cross-subject comparative analysis. For instance, Yeo's functional networks with rsfMRI [2] and Brainnetome atlas with dMRI [4] have been used to guide individualization based on K-means clustering. The process involves initializing the individual parcellation using a morphologically aligned reference parcellation, iteratively reassigning vertices to parcels based on the highest correlation between vertex timeseries and parcel reference signals, and updating the reference signals until convergence. The resulting individual functional networks

TABLE I
OVERVIEW OF OPTIMIZATION-BASED INDIVIDUAL BRAIN PARCELLATION METHODS

| Type | Key techniques | Year | Modality | Reference parcellation [a], number of parcels | Number of subjects | Surface-(S)/Voxel-(V) based | Objectives [b] |
|---|---|---|---|---|---|---|---|
| **Clustering** | K-means-based [2] | 2015 | rsfMRI | Yeo, 18 | Single | S | A (func) + B |
| | K-means-based [4] | 2020 | dMRI | Brainnetome, 210 | Single | V | A (anat) + B |
| | Local functional and structural connectivity fusion and N-cut clustering [9, 10] | 2015 | dMRI, rsfMRI | NA, 256/512/1024 | Single | V | A (func & anat) + C |
| | Simple linear iterative clustering (SLIC) [12] | 2016 | rsfMRI | NA, 50~1000 | Single | V | A (func) + C |
| | Joint K-means clustering [20] | 2016 | rsfMRI | NA, 100~200 | 2 | S | A (func) + B |
| | Multigraph K-way clustering [23] | 2013 | rsfMRI | NA, 100~300 | Multiple | V | A (func) + B |
| | Multi-view clustering [24] | 2016 | rsfMRI, sMRI | NA, 100 | Multiple | V | A (func & morph) + B |
| | Searching exemplar voxels and assigning left voxels to the exemplars [26] | 2018 | rsfMRI | NA, 2~25 | Multiple | V | A (func) + B |
| **Region growing** | Seeded region growing by fast marching [21, 27] | 2020 | rsfMRI | population-based parcellation [c], 200 ~ 4200 | Single | S | A (func) + B |
| | Atlas-guided parcellation (AGP) by seeded region growing [32] | 2022 | rsfMRI | multiple atlases [d], 200~1000 | Single | S | A (func) + B |
| | Seeded region growing and hierarchical clustering [14] | 2012 | rsfMRI | NA, 40~400 | Single | S | A (func) |
| **Template matching** | Match each vertex with template ROIs based on overlaps of their binarized RSFC maps [34, 35] | 2017 | rsfMRI | Yeo, 17 | Single | S | A (func) + B |
| | Match the patches separated from individual functional networks [2] with template regions based on overlaps [15] | 2019 | rsfMRI | Extended from Yeo's network, 116 | Single | S | A (func) + B |
| **Graph partition** | Subject-specific graph-based parcellation with shape priors (sGraSP), Markov random field [39] | 2017 | rsfMRI | GraSP, 76~2001 | Single | S | A (func) + B + C |
| | Individualize networks of subjects with brain lesions, Markov random field [42] | 2018 | rsfMRI | Yeo, 18 | Single | V | A (func) + B + C |
| | Group prior individual parcellation (GPIP), Markov random field [48] | 2017 | rsfMRI | Extended from Yeo's network, 58; HCP-MMP, 338[e] | Multiple | S | B (func) + C |
| | Community detection from RSFC using Infomap, and match communities with group networks based on overlaps [51] | 2017 | rsfMRI | population-based network [f], 17 | Single | S/V | A(func) + B |
| | Multimodal connectivity-based individual parcellation (MCIP) based on graph-cut [41] | 2024 | rsfMRI, dMRI | Brainnetome, 210; HCP-MMP, 360; Macaque brainnetome, 248 | Single | S | A (func) + B + C |
| **Matrix decomposition** | Non-negative matrix factorization [17] | 2017 | rsfMRI | NA, 17 | Multiple | V | A (func) + B + C + D |
| | Fuse connectivity matrix and spatial matrix, compute group-individual joint embedding, and obtain individual parcellation based on similarity of the joint embeddings [55] | 2023 | dMRI, sMRI | Macaque brainnetome, 248 | Multiple | S | A (anat & spat) + B |
| **Gradient-based** | Functional connectivity gradient and watershed segmentation [28, 29] | 2015 | rsfMRI | NA, variable | Single | S | A (func) + C |

NA, not applicable; ROI, region of interest; RSFC, resting state functional connectivity.

[a]The involved brain parcellation includes Yeo's functional networks [37], Brainnetome atlas [40], graph-based parcellation with shape priors (GraSP) [57], multi-modal parcellation (HCP-MMP) [3], and Macaque brainnetome atlas [58].

[b]The optimization objectives include three categories: intra-parcel signal homogeneity (A), inter-subject parcel homology (B), and spatial continuity (C). Another objective, pruning redundant parcels (D), was exclusively used in [17]. The intra-parcel homogeneity has been measured by functional (func), anatomical (anat), spatial (spat) and morphological (morph) features.

[c]The population-based parcellation is derived from multiple subjects in the study.

[d]Multiple parcellations have been used for validation including Shen 200 [23], Brainnetome 210 [40], Gordon 333 [29], HCP-MMP 360 [3], Schaefer 400 and Schaefer 1000 [47].

[e]HCP-MMP has 360 parcels, but the parcels less than 30 vertices are merged with their neighbors.

[f]The group-level network template is generated using Infomap from a group-average thresholded RSFC matrix.

demonstrate high reproducibility within subjects, capture variability across subjects, and have been validated through invasive cortical stimulation mapping in motor areas, highlighting their potential for clinical applications. Under the guidance of group-level reference parcellations, individualized functional networks not only capture subject-specific characteristics but also retain consistency across subjects. Similarly, the Brainnetome atlas has been employed as a priori in K-means-based iterative clustering using voxel-wise probabilistic tractography as the signals, giving rise to variations in individual parcellations [4].

In contrast to single-subject clustering, joint K-means, multi-graph/view, and exemplar-based clustering ensure inter-subject parcel homology by constraining the differences between individual parcellations across subjects. Joint K-means clustering has been applied to obtain homologous parcels for two resting-state sessions within a subject or between two subjects [20]. The objective function is designed to find two optimal clustering assignments by minimizing the functional distances between any vertex and its cluster centroid, as well as the difference between the two clustering assignments. In addition, multi-graph K-path spectral clustering utilizes features from multiple subjects and obtains group- and individual-level parcellations jointly based on rsfMRI [23]. This method optimizes intra-individual functional homogeneity within parcels while maintaining both group similarity and subject variability for each individual parcellation. Multi-view clustering with graph regularization has also been used to infer individual functional networks of multiple subjects based on sMRI and rsfMRI [24], which maintains inter-subject parcel homology by linking the parcellations of multiple subjects with a latent consensus parcellation. The graph regularization further encourages voxels with similar functional and morphological metrics to be assigned to the same network. Another approach to ensure inter-subject parcel homology is exemplar-based clustering [26]. In this method, a group of subjects is first used to select exemplar voxels that best represent each functional network. Formally, this is achieved by minimizing the sum of functional distances between the voxels of all subjects and the exemplars. The remaining voxels in each individual are then assigned to the nearest exemplar voxel based on functional similarity, resulting in individualized functional networks.

*2) Region growing*

Region growing methods begin by selecting “seed” voxels/vertices as starting positions for labeling, and then iteratively expand these seeds by incorporating neighboring voxels based on similarity or distance metrics until the entire brain/cortex is assigned to distinct regions. This process implicitly optimizes the similarity of features within parcels. Notably, the selection of seeds is crucial for individualization, as they are determined at initialization and cannot be updated with iterations. Some studies have designated the geometric center of each parcel in a reference parcellation as seeds, and other vertices are gradually allocated to these seeds to constitute parcels based on the similarity of resting-state functional connectivity (RSFC) between the vertices and the parcels [27, 32]. By using unified seeds from a reference parcellation, these methods maintain the homology of the same parcel across subjects. However, if the center vertex of the parcel chosen as the seed is affected by more noise, the resulting individual parcel becomes less reliable. A more robust solution is to select seeds based on functional stability [14]. First, stability maps are calculated as the root mean square error between all timeseries in a 3 mm neighborhood ROI and the ROI’s mean timeseries. Second, the local minima of the stability maps are identified as seed vertices. These seeds are then grown into patches, which are subsequently merged using hierarchical clustering to obtain the final individual parcellation. This approach has demonstrated high reproducibility within subjects, substantial overlap between individualized parcels and tfMRI-derived clusters, and clear RSFC transitions at parcel borders. However, this self-adaptive seed assignment method cannot generate homologous parcels across subjects.

*3) Template matching*

The template matching method initially selects a reference parcellation as the template and then matches voxels/vertices (or patches) on individuals to the most "similar" template parcel. This optimal matching between individuals and the template implicitly ensures both intra-parcel signal homogeneity and inter-subject parcel homology. Gordon and colleagues have used a similarity metric based on RSFC [34, 35], in which the top 5% values of each vertex's vertex-wise RSFC were binarized and compared with those derived from Yeo's 17 group-level functional networks. Each vertex is matched with a template network based on the overlap rate of binarized RSFC maps, resulting in 17 individual functional networks. Using this approach, individual parcellations exhibited consistency in their topographic arrangement with the group average template, but also contained unique topographic features not present in the template. The similarity metric could also be based on the topography of regions [15]. Individual functional networks were first derived from K-means-based clustering [2]. Subsequently, each network was separated into spatially separated patches and matched with one of the 116 regions in a template that demonstrated the highest overlap. It was shown that individualized regions increased the statistical power of tfMRI analyses, and the functional connectivity among these regions better predicted fluid intelligence compared to group-level parcellation. However, it cannot be guaranteed that all regions in the template can be detected in each individual.

*4) Graph partition*

A graph is defined as a set of nodes and a set of edges, each connecting two nodes. It is applicable to both surface-based and voxel-based representations of the brain. The cortical surface mesh, composed of triangular vertices and edges, is a type of graph. In voxel-based representations, each voxel can be viewed as a node, with the neighborhood of voxels defined as edges. Therefore, the individual parcellation problem can be solved by graph partition algorithms such as Markov random field (MRF) and graph-cut. Specifically, MRF is a graph partition algorithm capable of setting priors between neighboring voxels (or vertices), making it suitable for optimizing assignment similarity between neighboring points to enhance spatial continuity [39, 42, 48]. MRF is suitable for both single and multiple subjects. For single-subject optimization, subject-specific graph-based parcellation with shape priors (sGraSP) was proposed to derive individual-specific functional parcellations by optimizing an energy function consisting of intra-parcel rsfMRI signal homogeneity, reference parcellation similarity, and parcel spatial continuity [39]. Nandakumar et al. developed a method to derive individual functional networks with brain lesions based on rsfMRI [42]. In this approach, the individual network is initialized using Yeo's functional networks, and the voxel labels are iteratively updated by maximizing the correlation between each voxel's timeseries and its network's average timeseries while ensuring spatial continuity using MRF prior. To jointly obtain individual parcellations for a group of subjects, group prior individual parcellation (GPIP) initializes the parcellation using a reference atlas and, during the iterative optimization process, considers the RSFC similarity between individuals as well as the spatial continuity of parcels [48]. GPIP-derived parcellations exhibit distinct individual differences across subjects, show improved functional homogeneity within parcels, reduced variation in functional connectivity across subjects, and consistent individual parcel boundaries with regions identified based on tfMRI.

In addition to MRF, the Infomap and graph-cut algorithms are commonly utilized. Infomap is a community detection algorithm based on random walks on graphs and has been employed to accurately map brain functional networks at the individual level [51]. The process involves computing vertex/voxel-wise RSFC for each participant, thresholding it to retain the most relevant connections (0.3% to 5%), and using these thresholded matrices as input for Infomap to independently determine community assignments. The identified communities are then matched to a group-level network template generated using Infomap from a group-average thresholded RSFC matrix. The individual network assignment is obtained by aggregating assignments across thresholds, assigning each vertex/voxel to the labeled network corresponding to the smallest threshold. Multimodal connectivity-based individual parcellation (MCIP) uses graph-cut to maximize an energy function that optimizes both rsfMRI-derived functional and dMRI-derived anatomical homogeneity within each parcel of a subject [41]. The energy function also constrains similarity between the individual and reference parcellation while maintaining spatial continuity within each parcel. MCIP demonstrates good test-retest reliability, anatomical and functional homogeneity, the ability to predict behavior, and heritability.

*5) Matrix decomposition*

Matrix decomposition is commonly used to determine functional networks by decomposing fMRI timeseries into functional network components and their corresponding timeseries, which is a way to enhance signal homogeneity within a network. Common individual-level decomposition methods, such as back-reconstruction [59], dual regression [60], and group information guided ICA [61], are built upon independent component analysis and project group-level independent components onto individuals. However, these methods generate functional networks with negative values, making it difficult to interpret the biological significance of these negative areas. Therefore, non-negative matrix factorization (NMF) was used to jointly optimize non-negative individual functional networks for a group of participants [17]. The proposed method enhances functional consistency within networks by non-negative decomposition, ensures similar spatial distributions of the same functional networks across individuals using a group sparsity regularization term, encourages spatial smoothness by a data locality regularization term, and prunes redundant networks using a parsimonious regularization term. It is worth noting that this method focuses on obtaining individual functional networks with continuous values (i.e., loadings), and to convert them into parcels, each point needs to be assigned to the network with the highest loading (winner-takes-all rule). The obtained subject-specific functional networks exhibit improved resting-state functional homogeneity and coherence in task-related activations. In addition to fMRI-based decomposition, Laplacian decomposition is used to compute joint embeddings for a dMRI and sMRI fused group-average and individual connectivity matrix [55]. The connectivity matrix is fused by dMRI-derived anatomical connectivity and sMRI-derived spatial connectivity using similarity network fusion. After obtaining joint embeddings, each vertex is assigned to the parcel according to the joint embeddings. In this method, Laplacian decomposition enhances anatomical homogeneity, and the joint embedding between individual and reference connectivity matrices ensures homology of parcels across subjects.

*6) Gradient-based methods*

The gradient of RSFC for each vertex measures the extent of RSFC change compared to its neighboring vertices, which can be used for individual parcellation [28, 29]. Initially, the gradient of RSFC for all vertices is calculated. The resulting gradient map delineates the boundaries of regions with drastic RSFC changes, and individual parcellation is subsequently segmented using the watershed algorithm. Metaphorically, regions with slow RSFC changes resemble valleys that gradually get submerged by rising floodwaters, while ridges with drastic RSFC changes are not flooded and are identified as the boundaries of parcels. This approach implicitly ensures homogeneity within parcels, as vertices with slowly changing RSFC are assigned to the same parcel, as well as spatial continuality. By employing this approach, a reproducible subject-specific parcellation is obtained for an individual, which corresponds to subject-specific task activations. Additionally, the individual-level parcellation exhibits distinct topographic features compared to the group-level parcellation, although they share broad similarities. However, this method cannot ensure homologous parcels across subjects.

## *B. Learning-based methods*

Learning-based methods are characterized by estimating a mapping from individual-specific information to individual parcellation. These methods consist of two stages: training and inference. During the training stage, the learning model is constructed by minimizing a predefined loss function using individual features extracted from MRIs. At the inference stage, the trained model maps individual input features to individual parcellation. Formally, learning-based methods are expressed as:

$$\underset{W}{\operatorname{argmin}}\, L(\{X_i, P_i, G_i\}_{i=1}^{N}),$$
$$P_i = \phi_W(X_i), \quad (2)$$

where $\phi_W$ denotes the model (i.e., mapping), $W$ denotes the parameters of the model, $L(\cdot)$ denotes the loss function, $N$ denotes the number of subjects, $i$ denotes the index of a subject, $X_i$ denotes the individual-specific information, $P_i$ denotes individual parcellation, and $G_i$ denotes ground truth parcellation. Note that for some methods, $X_i$ or $G_i$ is optional in $L(\cdot)$. In contrast to optimization-based methods, learning-based methods can capture nonlinear relationships between individual-specific information $X_i$ and individual parcellation $P_i$ via the estimated mapping $\phi_W$. Because learning-based approaches generally require ground truth labels for model training, it is indispensable to determine the "true" individual parcellation as the ground truth $G_i$. However, a direct, accurate, and objective in vivo identification of whole-brain/cortex ground truth labels for individual regionalization is still lacking. Additionally, the brain is considered a hierarchical system, with various granularity in parcellation from fine-grained subregional specializations to the organization of functional networks having their own biological plausibility [62]. To our knowledge, three strategies have been employed to address the ground truth problem. Some studies directly employ a reference parcellation as ground truth [1, 3, 7, 11, 13], other studies use individual parcellation derived from other methods as ground truth [21, 22], while some studies introduce unsupervised strategies [16, 18, 25, 30, 31] (TABLE II). For all learning-based methods, a well-established model needs to be trained using sufficient participants.

Voxel/vertex-wise functional and anatomical connectivity derived from rsfMRI and dMRI results in high-dimensional data. To mitigate the risk of overfitting and enhance model performance, some learning-based methods extract features from functional networks [1, 3, 7, 21], parcel-wise functional connectivity [13], or tract-wise anatomical connectivity [11, 25] for dimensionality reduction. However, dimensional reduction may result in the loss of subtle information necessary for deriving individual-specific parcellations. Therefore, some models have attempted to deal with high-dimensional rsfMRI timeseries as input features, such as MS-HBM [16, 31], self-supervised CNN [18], and RefineNet [30].

### 1) *Using reference parcellations as ground truth*

Registering a reference parcellation from standard space to individual space provides a rough brain mapping at an individual level. However, the registration-based individual parcellation is a noisy ground truth, as a fraction of voxels/vertices in each region may not be correctly labeled. Therefore, the models should be robust enough to mitigate the effects of these mislabels. In practice, the models are either insensitive to the spatial position of the brain or supervised on constraints such as intra-parcel signal consistency and spatial continuity of parcels in addition to noisy ground truth.

Multilayer perceptron (MLP) and graph convolutional network (GCN) can derive labels based on learning the majority feature distribution of each parcel, benefiting from the insensitivity to the spatial location of each voxel/vertex. Specifically, MLP is an artificial neural network that consists

TABLE II
OVERVIEW OF LEARNING-BASED INDIVIDUAL PARCELLATION METHODS

| Type | Key techniques [a] | Year | Modality | Reference parcellation [b], number of parcels | Surface- (S)/Voxel- (V) based | Objectives [c] |
|---|---|---|---|---|---|---|
| **Using reference parcellations as ground truth** | MLP, multi-class classification [1] | 2013 | rsfMRI | Yeo, 7 | V | B |
| | MLP, binary classification [3] | 2016 | rsfMRI, tfMRI, sMRI | HCP-MMP, 360 | S | B |
| | MLP, binary classification [7] | 2022 | rsfMRI | MBMv4, 192 | S | B |
| | GCN, segmentation [11] | 2022 | dMRI | Brainnetome, 210 | S | B |
| | GCN, segmentation [13] | 2021 | rsfMRI | Schaefer, 400 | S | B |
| **Using individual parcellations as ground truth** | Integrating multiple geometric deep networks, segmentation [21] | 2021 | rsfMRI, tfMRI, sMRI | Individualized HCP-MMP, 360 | S | B |
| | Spherical CNN, segmentation [22] | 2022 | sMRI | Individualized MS-HBM, 400 | S | B |
| **Introducing unsupervised constraints** | Atlas individualizing with task contrasts synthesis (TS-AI), Spherical CNN, segmentation [25] | 2024 | rsfMRI, dMRI, sMRI, tfMRI (only for training) | HCP-MMP, 360; Brainnetome, 210 | S | A (func, anat & task) + B |
| | RefineNet, general neural network module [30] | 2022 | rsfMRI | Brainnetome, 246; Craddock, 200; AAL, 90 | V | A (func) + B + C |
| | Multi-session hierarchical Bayesian model (MS-HBM) [16, 31] | 2019 | rsfMRI | Yeo, 17 | S | A (func) + B + C |
| | CNN, segmentation [18] | 2023 | rsfMRI | NA, 17 | V | A (func) + B |

CNN, convolutional neural network; GCN, graph convolutional network; MLP, multilayer perceptron; NA, not applicable/available.

[a]Classification represents the model inputs the features of each voxel/vertex and outputs the voxel/vertex's the probability for each label. Segmentation represents the model inputs whole-brain/cortex features of each individual and outputs the individual's voxel/vertex-wise probability map for each label.

[b]The reference parcellation includes Yeo's functional networks [37], multi-modal parcellation (HCP-MMP) [3], MBMv4, Marmoset Brain Mapping Atlas version 4 [7], Brainnetome atlas [40], Schaefer's parcellation [47], Craddock atlas [50], and automated anatomical labelling (AAL) atlas [52].

[c]The optimization objectives include three categories: intra-parcel signal homogeneity (A), inter-subject parcel homology (B), and spatial continuity (C). The intra-parcel homogeneity has been measured by functional (func), anatomical (anat) and task features.

of fully connected neurons with nonlinear activation functions. Because of the shallow architecture, it has fewer parameters compared to deep networks [1, 3]. An individualization method used MLP to identify each voxel into one of the seven functional networks [1]. The voxel-wise RSFC features were dimensionally reduced using principal component analysis. This study selected voxels from a meta-analysis of task foci as the ground truth labels for training rather than using the entire group-level network as the ground truth to ensure that the selected voxels are accurately labeled. This multi-class classification on whole-brain voxels is challenging to extend to fine-grained parcellations because as the number of classes increases, the difficulty of the multi-class classification problem grows. In contrast, an MLP-based method proposed by Glasser et al. divided the multi-class classification task for all cortical parcels into multiple binary classification problems for each parcel [3]. The ground truth label for each MLP was set as the corresponding parcel in the reference parcellation. An MLP was trained for each parcel to determine if the vertex, which is within a 30 mm radius of the corresponding reference parcel, belonged to that parcel. For each vertex, features were extracted from sMRI, rsfMRI, and tfMRI, including myelination, cortical thickness, functional networks, task activation, and visuotopic maps. The binary classification results for each parcel were combined using the winner-takes-all rule, resulting in an individual parcellation of the entire cortex. The authors found that this approach delineates both typical and atypical parcel topography in individual subjects that cannot be identified by aligning group average parcellation with individual images using registration. This study suggests that different modalities provide complementary and essential information for individual parcellation, which is crucial for capturing topological deviations in individual subjects. A similar approach has been used on the marmoset cortex [7], while individual functional connectivity was used for training the MLPs instead of multimodal features.

GCN is another position-insensitive model that has been utilized for individual parcellation, where the triangular mesh of the cortical surface is treated as a graph, and vertex-by-tract anatomical connectivity [11] or vertex-by-parcel functional connectivity [13] is treated as node features. For instance, a two-layer GCN with Chebyshev graph convolution named Brain Atlas Individualization Network (BAI-Net) has been proposed [11]. A confidence-weighted cross-entropy loss is used to encourage consistency between the individual and reference parcellation. The loss ensures that the central area of a parcel is more decisive for guiding the output individual parcellation, making the ground truth more robust. BAI-Net provided reliable parcellations across multiple sessions and scanners within subjects. The topography of BAI-Net-derived parcellations showed better test-retest reliability and stronger associations with both individual cognitive behaviors and genetics, compared to the K-means-based individualization method [4].

*2) Using individual parcellations as ground truth*

Glasser's individual cortical parcellation [3] and MS-HBM-individualized parcellation [16] have been used as proxy ground truth to train individual parcellation models. A framework that integrates multiple geometric deep learning networks (gDL) has been proposed to obtain individual cortical parcellation with Glasser's individualized parcellation as proxy targets [21]. The model takes multimodal features derived from sMRI, rsfMRI, and tfMRI as input. The model leverages the vertex-to-vertex connections of gDL to improve the spatial continuity of the proxy targets. Similarly, by hypothesizing that individual functional parcellation can be inferred from more easily acquired sMRI (T1w and T2w), a spherical CNN is used to individualize functional networks from morphological features such as curvature, thickness, and myelination [22]. The ground truth functional networks of each subject were proxied by MS-HBM-individualized parcellation [31].

*3) Introducing unsupervised constraints*

In addition to relying solely on supervised ground truth, either reference or individual parcellations, some studies have included constraints, such as intra-parcel signal homogeneity and spatial continuity of parcels. For instance, the atlas individualizing with task contrasts synthesis (TS-AI) is proposed, which introduces intra-parcel homogeneity of multi-modal features, including rsfMRI-derived functional networks, dMRI-derived anatomical connectivity profiles, and task contrast maps [25]. TS-AI employs a two-stage pipeline based on spherical CNN. TS-AI first synthesizes task contrast maps for each individual by leveraging tract-wise anatomical connectivity and resting-state networks. These synthesized maps, along with feature maps of tract-wise anatomical connectivity and resting-state networks, are then fed into a U-net to obtain individual parcellation. By synthesizing task contrast maps, TS-AI does not require the existence of resource-intensive task fMRI scans during its application. The individualized parcellations produced by TS-AI were validated using various reference atlases and independent datasets, demonstrating the generalizability of the method. In addition to intra-parcel homogeneity, spatial continuity is also constrained in some studies. RefineNet uses intra-parcel rsfMRI signal homogeneity by maximizing the similarity between each voxel and its corresponding parcel's timeseries and constrains the spatial continuity of parcels by allowing label propagation between neighboring voxels [30]. Notably, RefineNet can be integrated as a module into various neural network architectures for different downstream tasks, such as physiological index prediction, disease classification, and tumor segmentation. Furthermore, the multi-session hierarchical Bayesian model (MS-HBM) combines intra-parcel RSFC homogeneity, the similarity between individual and reference parcellations, and spatial continuity [16, 31]. MS-HBM generates individual functional networks for multiple subjects by employing a hierarchical Bayesian model that considers multiple sessions of rsfMRI data. Recognizing that previous network mappings could potentially confuse intra-subject variability with inter-subject differences, MS-HBM hypothesizes that inter-parcel, intra-subject, and inter-subject RSFC follow von Mises–Fisher probability distribution, with the parameters estimated from the training data. Moreover, the topography of parcels derived by MS-HBM allows for the prediction of personal phenotypes encompassing cognition, personality, and emotion, which potentially serve as a fingerprint of human behavior.

Without constrained by ground truth parcellations, a self-supervised deep learning model has been proposed via a loss function comprising a data fitting term and a spatial sparsity term [18]. The data fitting term ensures intra-parcel signal homogeneity based on matrix decomposition, which aims to decompose the voxel-wise rsfMRI timeseries into several functional networks and their corresponding timeseries. The spatial sparsity term encourages overlapping between each functional network across individuals. This model first employs a set of temporally shared CNNs to extract temporal-invariant feature maps from the rsfMRI timeseries and then transforms the feature maps into individual functional networks using a U-Net architecture. The model, trained on healthy adults, demonstrated high intra-network homogeneity and prediction performance for new individuals from different datasets, including healthy individuals spanning various age ranges and patients with schizophrenia, indicating its good generalizability.

### C. Optimization objectives

Both optimization-based and learning-based methods involve a quantitative criterion for finding an "optimal" parcellation. Optimization objectives (i.e. prior assumptions) usually include intra-parcel signal homogeneity, inter-subject parcel homology and spatial continuity, as shown in TABLE I and II. At least one objective should be included in the method for individual brain parcellation.

#### 1) Intra-parcel signal homogeneity

Intra-parcel signal homogeneity suggests that the signals within a parcel should be more similar compared to those in other parcels. The signals vary across different methods, including functional timeseries or functional connectivity with rsfMRI, structural connectivity with dMRI, as well as morphology (e.g., curvature and thickness) and myelination with sMRI. Intra-parcel homogeneity can be implicitly achieved for region growing [14, 27, 32] and template matching [15, 34, 35] methods, in which each vertex/voxel/ROI is assigned to its most similar parcel/cluster within the algorithm to maintain intra-parcel homogeneity. Similarly, clustering [2, 4, 9, 10, 12, 26] and community detection [51] methods assign vertices/voxels with similar signals into a parcel, and gradient-based methods [28] assign slow-changing vertices in RSFC (i.e., vertices with similar RSFC) to the same parcel, implicitly satisfying intra-parcel homogeneity. In contrast, some graph partition [39, 41, 42] and some learning-based [16, 25, 30, 31] methods explicitly define intra-parcel homogeneity in the objective function, as the correlation or distance between the signal of each voxel and other voxels within the parcel. In addition, some clustering [20, 23, 24], self-supervised learning [18] and matrix decomposition [17] methods involve decomposing the voxel-wise signal matrix into a parcel-by-voxel membership matrix and a parcel-wise signal matrix [17, 18, 23, 24], which has been proven to be similar to grouping voxels with similar signals [63]. For learning-based methods, incorporating this assumption helps mitigate issues arising from the lack of ground truth individual parcellation, allowing the use of more complex deep networks in generating individual parcellations.

#### 2) Inter-subject parcel homology

Inter-subject parcel homology suggests that individual parcellations across different subjects should share the same number of parcels and exhibit functional homology for corresponding parcels. This assumption can be achieved both with and without involving a reference (i.e., template) parcellation. For methods that involve a reference parcellation, a popular approach is to guide individual parcellations to be similar to the reference parcellation [1, 3, 11, 13, 16, 21, 22, 25, 31, 39, 41]. This approach ensures consistency in the number of parcels, and the sizes and positions of each parcel are relatively consistent across subjects, thus considered homologous. Some methods, such as template matching, region growing, and exemplar-based clustering, obtain homologous parcels by assigning each vertex/voxel to the most similar parcel in a template/reference parcellation [15, 26, 27, 32, 34, 35, 51, 55]. Other methods achieve this while simultaneously ensuring intra-parcel signal homogeneity by enhancing the similarity between the signal of each point and that of its reference parcel [2, 4]. Notably, some methods only initialize the individual parcellation to be a reference parcellation but do not involve the reference parcellation after initialization [30, 42].

For methods that do not involve a reference parcellation, they should incorporate multiple subjects. Homologous parcels can be obtained by confirming a common latent parcellation derived from multiple subjects and constraining the similarity between the latent parcellation and each individual parcellation [23, 24]. Moreover, it is also feasible to directly encourage individual parcellations of each subject to be similar to each other [17, 18, 20].

#### 3) Spatial continuity

The assumption of spatial continuity suggests that each parcel should be spatially connected, rather than consisting of several discrete patches or small points. This assumption can be achieved both explicitly and implicitly. A common explicit approach is to penalize spatially neighboring voxels (or connected vertices) assigned to different parcels. As a result, the boundary voxels/vertices tend to be fewer, ensuring the spatial continuity of parcels. This approach can be formulated by matrix multiplication [17], MRF [16, 31, 39, 42, 48], and graph-cut [41]. Moreover, sGraSP further explicitly defined a star-convex shape prior to ensure spatial continuity based on MRF by guaranteeing that any two vertices belonging to the same parcel can be connected [39]. Other methods encourage spatial continuity implicitly by clustering a local connectivity matrix whose connectivity is only between each voxel and its neighboring voxels [9, 10], by incorporating spatial distance into the distance measure to constrain the search space to the neighborhood of cluster centers [12], or by propagating labels between neighboring voxels for each iteration [30].

### D. Modality

Resting-state fMRI records the blood oxygen level-dependent signal, reflecting in-vivo functional activity with relatively high spatial and temporal resolution, and has proven effective in capturing individual functional organization. Compared to task fMRI, rsfMRI is easier to acquire since it only requires the subject to remain at rest during scanning. However, the reliability of measuring the correlation between two rsfMRI signals depends on the scanning time duration, with longer durations yielding more reliable measurements. Consequently, rsfMRI scans of varying durations may result in distinct individual parcellations for the same subject [28]. Moreover, the reliability varies across the cortex, with regions in the inferior temporal and orbitofrontal areas often displaying lower fMRI quality [64]. As a result, different parcels derived from rsfMRI-based individualization methods may exhibit varying levels of confidence in their correctness.

Some methods incorporate dMRI-based fiber tractography to estimate whole-brain structural connectivity for individual subjects. Although structural connectivity tends to offer better reliability than functional connectivity derived from rsfMRI, a significant challenge is that fiber tractography is susceptible to false positive and false negative connections, potentially caused by gyral bias, unreliable pathway trajectories, and the underestimation of long-distance connections, etc. [65, 66]. Consequently, the derived tractographic weights only account for approximately 60% of the true fiber connections [65].

Certain methods employ multiple modalities to achieve individual parcellation because integrating various modalities is believed to provide a more comprehensive description of brain function and structural organization compared to a single modality. Examples include combining dMRI and rsfMRI-based connectivity [9, 10], combining rsfMRI and sMRI [24], as well as combining sMRI, rsfMRI, and tfMRI [3]. The primary challenge lies in the intricate correlations between these modalities—they share some common representations while also possessing modality-specific information. Existing methods have not thoroughly explored the relationships and complementarity between multiple modalities, nor have they deeply investigated how different modalities impact individual parcellation. Addressing how to fully exploit the rich individual information within each modality and integrate both consistent and distinctive information across modalities for individual parcellation remains an unexplored aspect in current research.

## III. Validating Individual Parcellation

As it is challenging to identify the ground truth parcellation in vivo, individual parcellation performance is commonly evaluated using indirect metrics such as intra-subject reliability and inter-subject similarity, intra-parcel homogeneity and inter-parcel heterogeneity, as well as correlation with personal characteristics. In addition, electrical cortical stimulation coherence, alignment with anatomical structures, simulation-based accuracy, etc., have also been used in accessing individual parcellations. Evaluation metrics in representative studies are listed in Supplementary Table. Publicly available datasets enable consistent and fair comparison across algorithms [18, 35, 51, 67-74]. See detailed introduction of the roles these datasets played in individual parcellation in Supplementary Material.

### A. *Intra-subject reliability and inter-subject variability*

Individual partitions obtained at different scan sessions of the same subject should be similar, and thus intra-subject reliability (or robustness) can be evaluated in terms of the overlap rate (Dice coefficient) between intra-subject parcellation pairs. The Dice coefficient is defined as the ratio of points assigned to the same parcel in two MRI scans to all the points [2, 4, 9-14, 16, 20, 26, 27, 31, 32], which can be formulated as:

$$\text{Dice} = \frac{2 \times |A \cap B|}{|A| + |B|}, \quad (3)$$

where $A$ and $B$ denotes parcellations derived from two MRI scans within a subject, $A \cap B$ denotes points identically labeled in $A$ and $B$, and $|\cdot|$ denotes the number of points. In contrast to intra-subject reliability, individual parcellations of different subjects should be diverse, and thus inter-subject variability can be evaluated using the overlap rate of parcellations between subjects, which can also be formulated using Equation (3), where $A$ and $B$ denotes parcellations of two subjects.

It is worth noting that intra-subject reliability and inter-subject variability are usually evaluated in pairs, and a qualified individualization method should satisfy that intra-subject reliability is greater than inter-subject variability [16, 31]. Some studies have used the gap between intra-subject reliability and inter-subject variability as a metric, with a larger gap indicating better individual parcellation [2, 11, 13]. Another comprehensive metric for assessing intra-subject reliability and inter-subject similarity is the accuracy of individual identification. This metric involves computing the intra-subject and inter-subject overlap rates between each test-retest subject pair. Subsequently, each subject is assigned to the subject in the retest dataset with the highest overlap rate. The identification accuracy is calculated as the ratio of the number of participants correctly assigned to their respective identities to the total number of subjects [18, 27, 32], with higher accuracy indicating better performance.

### B. *Intra-parcel homogeneity and inter-parcel heterogeneity*

#### 1) *rsfMRI-based homogeneity*

The idea that signals (rsfMRI timeseries or RSFC) within the same parcel should be as consistent as possible, whereas signals in different parcels should differ, is a fundamental assumption of functional parcellation. Many studies have used functional homogeneity as a measure of the degree of intra-parcel signal agreement, usually calculated as the average correlation coefficients between all pairs of points within each parcel [9, 10, 12, 13, 18, 22, 31, 39, 42, 48], denoted as:

$$\text{Homo} = \frac{\sum_l \rho_l |l|}{\sum_l |l|}$$
$$\rho_l = \frac{\sum_{\substack{v,u \in l \\ v \neq u}} \text{corr}(\boldsymbol{x}_v, \boldsymbol{x}_u)}{|l|(|l|-1)/2}, \quad (4)$$

where $u$ and $v$ denotes two different points within parcel $l$, $\text{corr}(\boldsymbol{x}_v, \boldsymbol{x}_u)$ denotes Pearson's correlation coefficient between the functional signal of $v$ and that of $u$, and $|l|$ is the number of vertices within parcel $l$. Other studies calculate the average correlation coefficients between the timeseries of all points and their centroids within each parcel (the higher the better) [17], the percentage of variance explained by the first principal component of RSFC for all vertices within each parcel (the higher the better) [28], the distance between the timeseries of point pairs within each parcel (the lower the better) [24], or the entropy of parcel-wise timeseries measuring the amount of noise in BOLD signals (the lower the better) [39].

*2) dMRI-based homogeneity*

Intra-parcel homogeneity can also be defined based on anatomical connectivity profiles, where the signal $\boldsymbol{x}$ in Equation (4) is vertex-wise anatomical connectivity derived from fiber tractography based on dMRI [4], which is proposed to validate dMRI-based individual parcellation.

*3) tfMRI-based homogeneity*

Brain activation evoked by tasks can be recorded using tfMRI, which reveals subject-specific functional topography on the cortex. It is widely acknowledged that individual parcellation should be spatially correlated with task-evoked activation, meaning that highly activated regions should be located more within parcels instead of on parcel boundaries. Some studies visually inspect this coherence, such as language area 55b with language task contrast maps [3, 21] and early cortical visual areas V1, V2, and V3 with retinotopy maps [28]. Quantitatively, the variance of the task contrast map in each parcel is measured, termed as task inhomogeneity [13, 16, 31, 48, 51], with a lower value indicating more consistent task activation and smaller changes within parcels, formulated as:

$$\text{Inhomo} = \frac{\sum_l \text{std}_l |l|}{\sum_l |l|}, \tag{5}$$

where $\text{std}_l$ denotes the standard deviation of all point values within parcel $l$ of a task contrast map. Similarly, task homogeneity is defined as the inverse of the standard deviation of the task contrast map in each parcel [25, 41]. Additionally, the voxel-wise spatial correlation between functional networks and task activations [17] and the percentage of thresholded task activation within individual parcels [28] are also used, with higher percentages indicating greater coherence.

Some studies evaluate the temporal correspondence between individualized parcels and task activation. The signal consistency between the motor network and its corresponding task activation regions is defined as the correlation between the average timeseries of individual resting-state networks and those of highly activated regions during motor tasks [42]. Similarly, Li et al. defined the temporal correlation between functional networks and task activations [17]. Some studies compared individualized parcellation with tfMRI across subjects, where the similarity of parcellated task activation maps between different subjects evaluated the task activation coherence across subjects, with individualized parcellation expected to obtain higher similarity compared to group-wise parcellation [15]. The average activation of each parcel across subjects can also evaluate inter-subject task activation coherence [13].

*4) Combining homogeneity with heterogeneity*

While intra-parcel homogeneity only considers the consistency of signals within the same parcel, some studies consider both homogeneity within parcels and heterogeneity between parcels using metrics such as silhouette coefficient [27, 32] (the larger the better), Dunn index [24, 26] (the larger the better), or Davies-Bouldin index [26] (the smaller the better).

*5) Correlation with personal characteristics*

The variability of functional organization has been shown to correlate with various personal characteristics and brain disorders. These correlations can be used to examine the effectiveness of individualization methods. Previous studies showed that sex [26] and age [18] can be classified by functional network topography, and young/old can be classified by RSFC [17]. Fluid intelligence can be predicted based on RSFC among individualized regions [15, 30]. More behaviors that reveal a wide range of cognition and personality, e.g., reading (pronunciation), cognitive flexibility, working memory, etc., have been used to examine the correlation with individual parcellation-derived RSFC [31] and individual parcellation topography [11, 13, 16, 18]. Individuals with closer genetic relationships should exhibit more similar parcellations [11], indicating that the heritability encoded in the parcellation topography can evaluate individual parcellation methods. Individual parcellation also encodes information about brain diseases and disorders, e.g., autism spectrum disorder [30] and schizophrenia [18], as improved classification accuracy can be achieved using individual-specific RSFC compared to group-level parcellation-derived RSFC. Moreover, the accuracy of tumor ROI classification has been shown to improve using individual-specific parcellation [30].

### C. *Other validations*

**Electrical cortical stimulation (ECS) coherence**: Wang et al. [2] located hand and tongue sensorimotor regions using ECS, the "gold standard" for preoperative functional mapping. Their methods showed greater consistency with ECS compared to task MRI-defined regions, but ECS is invasive and resource-intensive.

**Alignment with anatomical structures**: Individual parcellation borders align well with changes in cortical myelination [10, 14], anatomical atlases [14], cytoarchitectonic atlases [10], and morphological features [1].

**Simulated data**: Simulating timeseries for multiple subjects can assess false positive and false negative rates [17, 18, 23, 48]. Importantly, inter-subject variability should be introduced during data generation, which may involve image translation, rotation, spread, and adding noise [17, 18].

**Parcellation consistency**: The group parcellation averaged from individual parcellations should resemble the group-level parcellation used as the ground truth [3]. The parcel detection rate is also introduced [3], which quantifies excessive shrinkage or expansion of individual parcels.

**Spatial discontinuity**: Measured by the number of spatially discrete components within each parcel [12, 31], smaller values indicate better spatial continuity.

## IV. APPLICATIONS

### A. *Individualization in neuroscience research*

Individual parcellation holds great promise for understanding functional and anatomical variations among individuals. Individual parcellations revealed the uniformed distribution of inter-subject variability across the cortical cortex, with higher variability in heteromodal association cortex and lower variability in unimodal cortices, by both rsfMRI-based [15, 16, 20, 31, 48, 75] and dMRI-based individualization methods [11].

The topography of individual parcellation is inherently linked to a wide range of individual characteristics, such as age, sex, and behavior. Several studies have demonstrated the potential of individual parcellation in predicting these characteristics. For instance, researchers have employed machine learning approaches, such as CNN and support vector machines (SVM), to predict brain age [76] and classify the sex of participants [77] based on the topography of individual functional networks. A multi-scale functional connectivity brain age prediction model, constructed using individual functional networks from a large-scale multi-site fMRI dataset, revealed significant correlations between brain age discrepancy scores and clinical and cognitive measures [78]. Individual-specific connectivity predicts cognitive behavioral scores more accurately than group-level approaches [31]. The co-representation in individualized functional networks aligns spatial coordination with cytoarchitectural classes, and has been found to be a reliable predictor of individual behavioral performance [79]. The Overlapping MultiNetwork Imaging (OMNI) generates group-level probabilistic networks that allow for overlapping across different networks, enhancing the reliability of behavior-RSFC correlations compared to group average-based parcellations [80]. Individual parcellation also helps in identifying network variations associated with different types of childhood maltreatment [81].

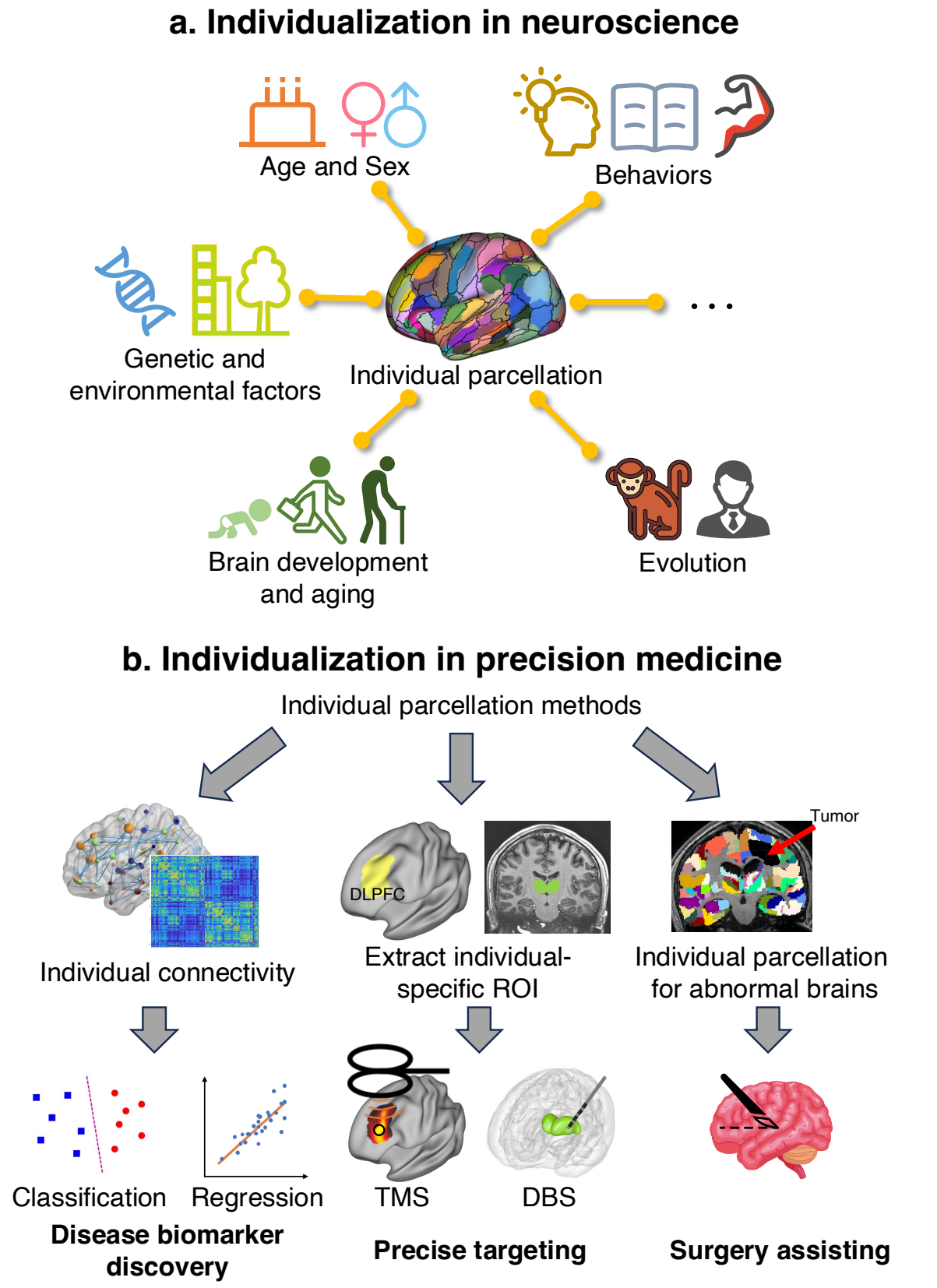

Fig. 2. **a.** Individual parcellation has significant implications for advancing neuroscience research. It has revealed associations between individual brain organization and factors such as age, sex, cognitive abilities, genetic influences, brain development and aging, etc. Individual parcellation has also facilitated cross-species investigations. **b.** Individual parcellation has found wide applications in identifying connectivity biomarkers for various disease symptoms, thereby aiding in the diagnosis of brain disorders. Methods have been developed to parcellate brains with structural abnormalities such as tumors, which assisted surgical procedures. Additionally, individual parcellation enables precise targeting of therapeutic regions in treatments like transcranial magnetic stimulation (TMS) and deep brain stimulation (DBS).

Studies across various age groups, from neonates to the elderly, have utilized individual parcellation to explore developmental and aging patterns in brain networks. In neonates, the most individual variability was found in the ventral attention networks [82]. During youth, functional network topography improves with age, encoding individual brain maturity and predicting individual differences in executive function [33, 83, 84]. In contrast to younger individuals, aging brains exhibit a pattern of de-differentiation, characterized by reduced modularization and increased isolation among different functional organizations. Studies using individual parcellation have revealed global and network-specific de-differentiation patterns in aging brains [85]. Furthermore, aging-related dynamic network changes have been characterized, showing a compensative increase in emotional networks and a decline in cognitive and primary sensory networks in healthy aging [86]. A method for generating age-specific group atlas has been proposed based on individualized parcellation, establishing the first set of age-specific brain atlases covering the entire lifespan [87].

Research also indicates genetic and environmental influences on individual functional network organization. Twin studies have shown that genetic factors affect the size and topology of cortical networks [38]. Furthermore, individual environments and experiences in childhood have been found to be associated with various cognitive functions and are reflected in the individual topography of functional network organization [88]. Additionally, differences in network topology have also been observed between Caucasian Americans and Han Chinese, highlighting racial diversity in brain organization [89].

Individual parcellation methods have been adapted to non-human primate animals, facilitating comparative and translational neuroscience research. In marmosets, an individual parcellation approach using rsfMRI and multiple binary classifiers has been introduced [7], similar to the method employed by Gasser et al. [3] in humans. For macaques, an individual parcellation method integrating dMRI-derived structural connectivity with sMRI-derived spatial connectivity has been proposed [55]. Cui et al. [41] computed the topographic variability of individual parcellations via MCIP in both humans and macaques, showing that the size and position of parcels in humans exhibit greater variation across individuals compared to those in macaques.

### B. *Individualization in precision medicine*

#### 1) *Disease biomarker discovery*

Individual brain parcellation captures subtle individual-specific morphology, connectivity, and brain activity characteristics, offering the potential to more accurately identify neuropsychiatric disorders and predict the severity of symptoms. Statistical analysis has been used to compare differences among groups and correlate them with symptoms. For instance, using ANOVA, a study compared functional network topologies across typical development, autism spectrum disorder (ASD), and attention-deficit/hyperactivity disorder (ADHD), finding distinct differences in network properties like small-worldness and clustering coefficients [90]. Another study linked individual topographic features with Parkinson's disease severity, using a region-growing algorithm to detect biomarkers related to disease progression [32].

In contrast to statistical analysis, machine learning techniques have attracted interest for disease diagnosis and prediction. The process typically involves extracting features from individualized brain parcellations, selecting optimal features, and training classification or regression models. Various features can be extracted, including functional connectivity among individual parcels/functional networks [43-46, 91-94], individual parcel/functional network topography [32, 95], and individual graph-theory-based properties [90]. Selecting the most representative features improves diagnostic accuracy and aids in understanding disease progression by identifying biomarkers [46, 93]. Machine learning models, such as SVM, support vector regression (SVR), and graph neural networks (GNN), are then trained and established to discriminate between patients and healthy participants or to correlate individual characteristics with symptoms.

Individual ROI-wise functional connectivity is widely utilized to distinguish patients from healthy participants, which has shown superior performance compared to those from group-level parcellation. For instance, using individual parcellation, SVM classifiers achieved an 83% accuracy rate, outperforming group-level parcellation's 74% [44]. Furthermore, dual-branch GNNs enhanced classification of cognitive impairment stages by incorporating both individual and group-level connectivity features [91].

In addition to differentiating patients from normal participants, functional connectivity [43, 45, 92-94] and parcel topography [34] have been used to predict symptom scores, with improved performance compared to group-level parcellation [45, 92, 94]. For instance, functional connectivity calculated from individual parcels that showed a high correlation with symptom scores was selected to train the SVR model, resulting in significantly improved regression performance with PANSS symptoms in patients with schizophrenia or schizoaffective disorder [45]. Individual parcellation-derived functional connectivity has also been used to track changes in symptom severity [46, 93]. Additionally, functional network topography obtained from NMF [17] predicted four-dimension factor psychopathological scores in youth using partial least squares regression, showing high predictive performance.

Interpretability analysis identifies novel biomarkers and enhances understanding of disease mechanisms, providing accurate evidence for diagnosis and treatment. Machine learning studies often identify features contributing to classification/regression by counting the frequency [91, 92] or weights [43, 93-95] of each feature selected across cross-validation folds, with higher values indicating greater contribution.

### *2) Neuromodulation*

Individual-specific parcellations enable precise targeting of therapeutic regions in non-invasive treatments like transcranial magnetic stimulation (TMS) and invasive treatments such as deep brain stimulation (DBS). Individual parcellation and functional connectivity-driven methods have been proposed to determine TMS target regions, showing promising results in patients with various neurological conditions, including major depressive disorder [96], postoperative neurocognitive deficits following brain tumor resection [97], and motor or language deficits after neuroglial tumor resection [98]. Targeted functional network stimulation (TANS), using individual-specific functional connectivity maps, optimizes TMS procedures, improving motor outcomes and reducing side effects in Parkinson's disease patients [99]. Comparative analyses between group-level and individual functional network-based TMS targeting have highlighted the advantage of individual targets in ensuring the reliability of TMS responses during each treatment session [100]. In DBS, an individualized parcellation scheme for the subthalamic nucleus has been proposed, segmenting it into three subregions for each participant [101]. The overlap of these subregions with effective DBS treatment areas demonstrates the feasibility of this approach in clinical settings. Further applications and details of precise targeting in TMS and DBS can be found in another review [49].

### *3) Surgery assisting*

Accurate delineation of individual parcellation in brains with anatomical abnormalities resulting from injuries facilitates surgical procedures and enhances understanding of pathogenic mechanisms. A dMRI-based approach was proposed to establish individual parcellations for brains with deformations caused by brain diseases or resection surgery [19]. This method derives structural connectivity from normal human and trains a model to classify each voxel based on this connectivity. An individual network parcellation technique for brain tumor patients was developed using rsfMRI [102], leveraging Wang et al.'s [2] individualization method. Results revealed that 33.2% of brain tumors were intra-tumoral functional areas with primary sensory and higher cognitive functions. A method for localizing individual functional networks with abnormalities in brain glioma patients was proposed [103], showing that gliomas cause wide changes in network topography, mainly in structurally normal regions outside the tumor and oedema region.

## V. Prospects and Conclusions

In our review, we classified individual parcellation methods into optimization-based and learning-based categories. Optimization-based methods involve traditional techniques that were originally explored for individual brain mapping. They rely on prior assumptions and can be applied to individual information derived from single or multiple subjects. However, these methods may not capture high-order and nonlinear correlations between the individual-specific information and the individual parcellation. In contrast, learning-based methods require a number of participants for model training, and thus high-order and nonlinear correlations between the individual-specific information and the individual parcellation can be captured. It is evident that these methods have the potential to achieve superior performance if the models are well-established [11, 18], as they can capture more complex relationships between individual features and individual parcellation. In this sense, optimization-based methods are more applicable in scenarios where acquiring a sufficient number of subjects is challenging, such as individual parcellation for patients with brain diseases and nonhuman primates. For learning-based methods, the introduction of additional constraints, such as intra-parcel homogeneity and spatial continuity, has been demonstrated to effectively mitigate problems caused by the lack of ground truth parcellation. However, the generalizability of learning-based methods poses a challenge for new subjects from independent datasets, especially for individuals with brain dysfunction that differ from those in the training set. Additionally, learning-based methods require the extraction of the same features as the training data, thus limiting the generalizability of individual parcellation models to subjects from independent datasets with missing modalities that are difficult to acquire or have low MRI imaging quality. Instead of MRIs, more in-vivo signals, such as electroencephalography and electrophysiology, providing different spatial and temporal resolutions, can also reflect the individual variability of the functional and anatomical organization of the brain, which may usable for defining individual parcellations with MRIs or independently. Recent advances in fundamental models pretrained on large data have shown to be adapted to a wide range of downstream tasks. It is anticipated that individual parcellation will rapidly evolve in the direction of advanced artificial intelligence approaches such as large-scale neuroimaging-based models [104-106]. In addition, designing a generalizable individual mapping tool is highly desirable, models that generalizable to multi-site and multi-subject data will reduce the gap between individual parcellation research and its use in actual clinical practice.

The lack of ground truth parcellation in individuals poses a common challenge in validation. Most studies have validated their proposed methods using intra-subject reliability, inter-subject similarity, intra-parcel homogeneity, and inter-parcel heterogeneity. Some studies have introduced external data, such as task-evoked activation, anatomical structures, and personal characteristics, to assess the coherence or correlation of parcel topography with these factors. A potential advance for validation is to explore more in-vivo and ex-vivo measurements as validation tools. ECS is an in-vivo measurements that was used by Wang et al. [2] to validate the precise location of hand and tongue sensorimotor regions. However, clinical data is rare and challenging to acquire. Other in-vivo measurements, such as electroencephalography and electrophysiology, may also be suitable for validating individual parcellations, but to our knowledge, no investigations have involved these so far. Ex-vivo cytoarchitectonic histology, such as Nissl-stained images, reflect anatomical organization at a high-resolution level and can thus be tools for validating individual parcellation. However, due to the complexities of acquiring and processing histological data, current studies have not extensively explored this avenue. Publicly available datasets that incorporate these new paradigms in validation, such as the histology datasets BigBrain [107] for humans and BigMac [108] for macaques, would be invaluable for advancing the field.

Individual parcellations have the potential to facilitate neuroscience research and precision medicine, while their application in research and clinical environments can verify the accuracy and effectiveness of individual brain mapping. The topography of individual parcels, associated with multiple personal characteristics affected by genetic and environmental factors, encodes information on brain development and evolution. Biomarkers of various neurological and psychiatric diseases/disorders have been identified based on individual parcellations, emphasizing the importance of personal brain organization. However, most current studies only consider a single individualization method, despite the potential for variant performance and biomarker identification across methods. Fair comparisons across multiple methods are crucial for evaluation and robust biomarker identification. The application of individual parcellation in brains with anatomical abnormalities holds great clinical value but requires broader innovations, as existing methods predominantly focus on normal participants. While individual parcellation has been integrated into some TMS and DBS protocols for precise targeting, larger-scale, multi-disease studies are needed to systematically evaluate their efficiency compared to traditional targeting methods.

Despite the clear demonstrations of advancing our understanding of human brain organization and the potential to generate novel diagnostics and therapeutics for brain diseases, the field of individual brain mapping is still developing. Numerous unique challenges related to establishing ground truth for supervised models and validations, addressing patients with brain dysfunctions and verifying the validity of individual mapping are still being tackled. The establishment of an open science resource with publicly accessible data, tools and evaluation techniques that will facilitate the development of individual parcellation mapping and its applications for healthy, abnormal, developmental and aging brains. Collaboration and open science within the individual parcellation community will encourage more investigators in the fields of neuroscience, artificial intelligence and clinical medicine to contribute to accurate brain mapping at an individual level. The code and HCP subjects' individual parcellations based on multiple atlases derived from MCIP and TS-AI, proposed by us, are currently available on in https://github.com/YueCui-Labs. We plan to provide more resulting data and develop user-friendly toolkits in the future.

In this review, we attempted to conduct a comprehensive survey on the progress of innovative and effective individual brain parcellation methods, validations, and applications in the literature. We summarized these methods based on algorithms, optimization objectives, modalities and datasets, as well as the metrics for validations. Broad applications across neuroscience and clinical research have emphasized the great importance of individual parcellation. We appeal for an open platform integrating multifaced resources including multimodal and multiscale datasets, algorithms, tools and evaluation techniques, paving the way for the development, enhancement and convenient utilization of individual brain mapping.